\documentclass{iopart}
\usepackage{iopams}
\usepackage{graphicx}
\usepackage{color}
\usepackage{ulem}
\usepackage{listings}

\begin{document}
\title[Thermal Correction of Astigmatism in GEO\,600]{Thermal Correction of Astigmatism in the Gravitational Wave Observatory GEO\,600}
\author{H\,Wittel$^{1}$, H\,L\"{u}ck$^{1}$, C\,Affeldt$^{1}$, K\,L\,Dooley$^{1}$, H\,Grote$^{1}$, J\,R\,Leong$^{1}$, M\,Prijatelj$^{1}$, E\,Schreiber$^{1}$, J\,Slutsky$^{1}$, K\,Strain$^{2}$ M\,Was$^{1}$, B\,Willke$^{1}$ and K\,Danzmann$^{1}$}
\ead{Holger.Wittel@aei.mpg.de}
\vskip 1mm
\address{$^{1}$\,Max Planck Institute for Gravitational Physics and University of Hannover, D-30167 Hannover, Germany}
\vskip 1mm
\address{$^{2}$\,SUPA, School of Physics and Astronomy, The University of Glasgow, G12\,8QQ, United Kingdom}


\begin{abstract}
The output port of GEO\,600 is dominated by unwanted high order modes (HOMs). 
The current thermal actuation system, a ring heater behind one of the folding mirrors, causes a significant amount of astigmatism, which produces HOMs. 
We have built and installed an astigmatism correction system, based on heating this folding mirror at the sides (laterally). With these side heaters and the ring heater behind the mirror, it is possible to tune its radius of curvature in the horizontal and the vertical degree of freedom. We use this system to match the mirrors in the two arms of GEO\,600 to each other, thereby reducing the contrast defect. The use of the side heaters reduces the power of the HOMs at the output of GEO\,600 by approximately 37\%.
\end{abstract}

\section{Introduction}
The gravitational wave (GW) observatory GEO\,600 is a 600\,m long dual-recycled \cite{dual} Michelson interferometer with folded arms. GEO\,600 is operated close to the dark fringe, such that most of the light is reflected towards the input port. The ouput port is therefore called the dark port. For the dual recycled operation, GEO\,600 has two mirrors added to the basic Michelson topology. A highly reflecting mirror at the input port forms a high finesse cavity, called the power recycling cavity (PRC). The other recycling mirror is located at the output port of the interferometer and forms the signal recycling cavity. Mirror imperfections convert light from the TEM$_{00}$ mode to higher order modes (HOMs), which are not resonant in the PRC and leave the interferometer. The power at the dark port of GEO\,600 is dominated by HOMs, which do not carry a useful gravitational wave signal for GEO\,600, and only contribute to shot noise. The TEM$_{00}$ light at the darkport is composed of about 6\,mW of GW signal carrier light for the DC readout \cite{DC}, and RF sidebands to this carrier light. These sidebands are used for controlling the interferometer. Their power is very low compared to the carrier, thus they will be neglected for this work. A small cavity, the output mode cleaner (OMC) \cite{omc}, is installed in front of the main photodiode to filter out the HOMs and RF sidebands. HOMs can cause problems, however. When the OMC is not optimally aligned, HOMs can overlapp \cite{mirko} with the TEM$_{00}$ eigenmode of the OMC and decrease the signal to noise ratio for gravitational waves on the main DC readout photodiode. 
In the signal recycling cavity HOMs are partially converted back to the TEM$_{00}$ mode \cite{dual}. In the case of GEO\,600, it has been observed that this process depends strongly on the alignment of the signal recycling mirror. Therefore, in the presence of HOMs, small fluctuations of the signal recycling cavity can lead to power fluctuations in the interferometer. The largest source of HOMs in GEO\,600 is astigmatism.

\begin{figure}
	\centering
		\includegraphics[width=1.00\textwidth]{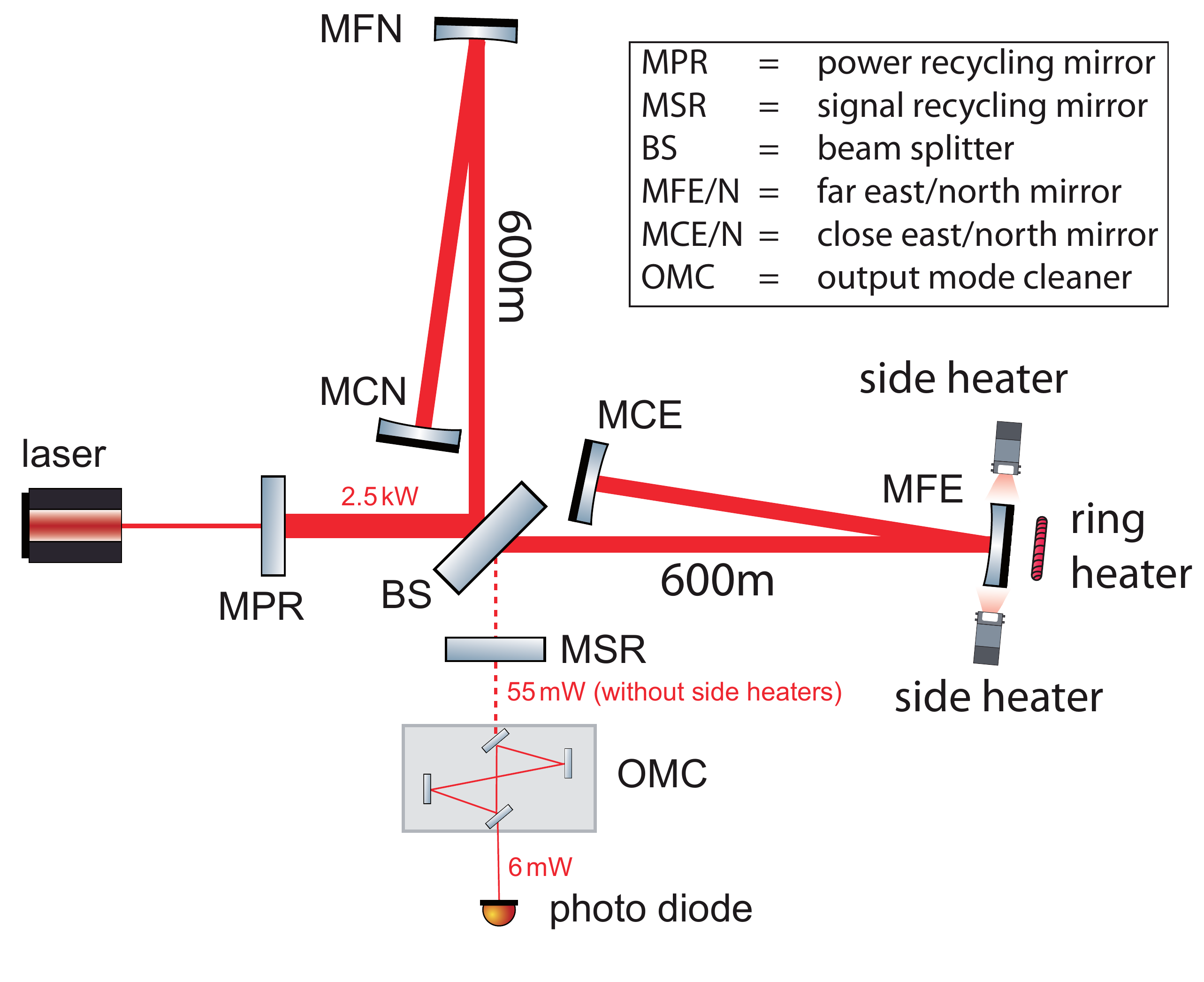}
	\caption{A simplified layout of GEO\,600. The power of the laser field at different points in the interferometer is given in red numbers.}
	\label{fig:GEO-layout-neu-simple-paper}
\end{figure}

\section{The Ring Heater and Astigmatism}
All large scale gravitational wave detectors require very high quality optics. Since the specifications for the optics are very demanding, some mirrors deviate from the specified radius of curvature (RoC).
Both GEO\,600 (\cite{heater}, \cite{StefanDiplom}) and recently Virgo (\cite{chrocc}), the Italian-French GW observatory, have developed and used systems to correct the curvature of a mirror in one degree of freedom by application of thermal radiation. Ring heaters are also planned to be used in advanced LIGO  \cite{aligo}. Additionally, LIGO and Virgo have also used CO$_2$ laser based heating \cite{ballmer}, \cite{accar}.

GEO\,600 has been using a ring heater to match the RoC of the far east mirror to the RoC of the far north mirror since 2004. The ring heater sits behind the mirror. It consists of a metal ribbon that is wrapped around a glass ring. When a current is sent through the metal ribbon, its temperature increases. The thermal radiation from the ribbon is absorbed by the mirror, and due to the thermal gradient from back to front, the mirror deforms and changes its RoC. Without the ring heater, we cannot lock the PRC of GEO\,600.

We have studied a finite elements model of the ring heater setup and found that it causes a significant astigmatism in the far east mirror. This is due to the fact that the thermal radiation of the ring heater is not completely directed towards the mirror, but also reaches structures underneath the mirror, which in turn heat the bottom of the mirror.
By matching the measured beam shape at the dark port to a simulated one (using the software FINESSE \cite{finesse}) we find that the dark port beam shape can be explained by a RoC of the far east mirror of 665.5($\pm$2)\,m in the horizontal direction and 658($\pm$2)\,m in the vertical direction. This was done with the ring heater at the setting that gave the lowest power at the dark port.

Ideally, the curvature would be identical in both directions. We decided to design and install additional heaters at the sides of the far east mirror which would cause a local thermal expansion of the mirror at the sides and as a result, decrease the RoC in the horizontal direction. This way we can adjust the mirror RoC in two degrees of freedom (horizontal and vertical). A simplified layout of GEO\,600 including the heaters is shown in fig.\ref{fig:GEO-layout-neu-simple-paper}.

\section{Side Heater Design}
\begin{figure}
	\centering
		\includegraphics[width=1.00\textwidth]{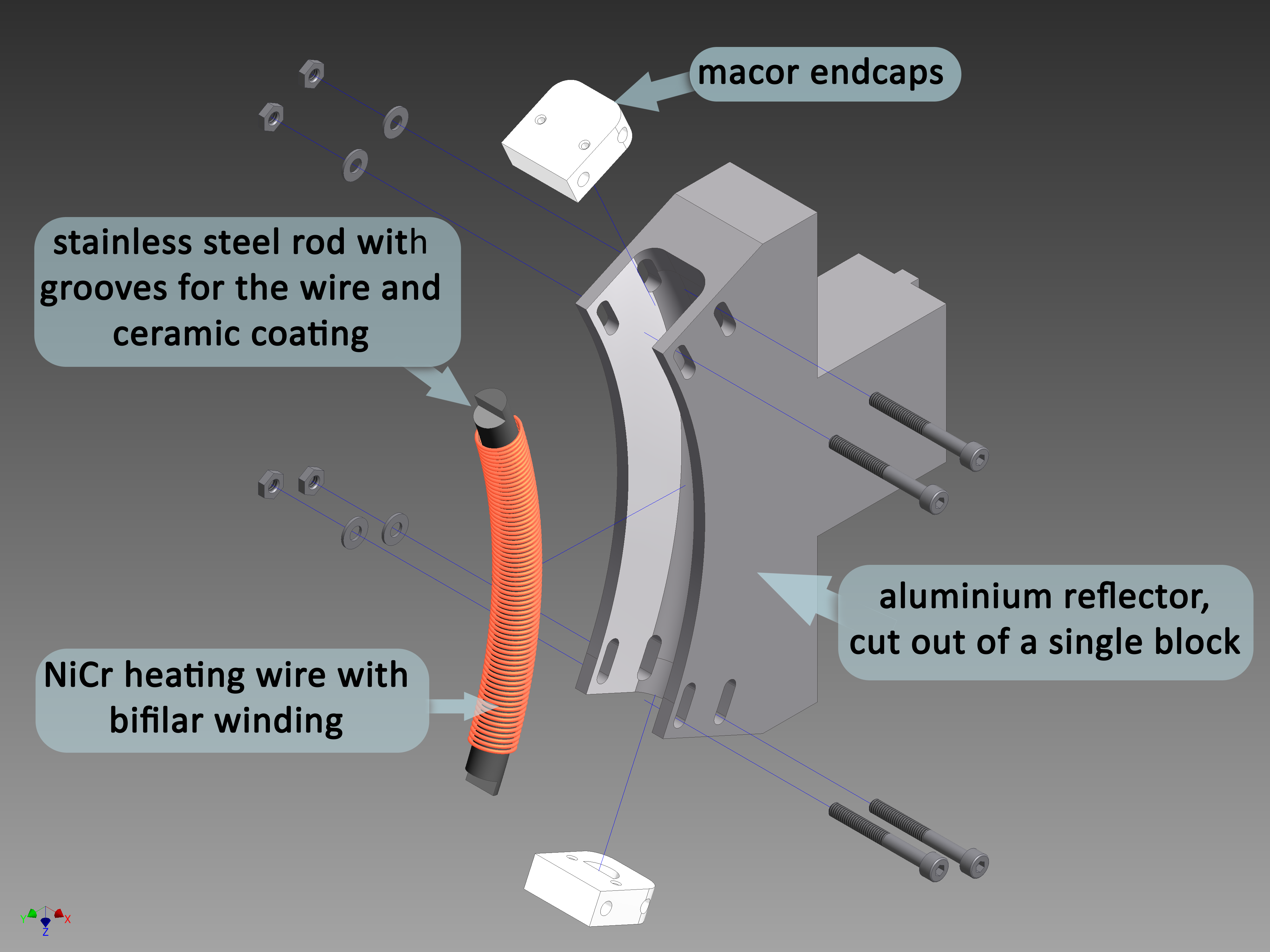}
	\caption{CAD drawing of one of the side heaters}
	\label{fig:Assembly4-only-for-presentation-with-labels}
\end{figure}

Each of the two side heaters consists of a curved, ceramic-coated, stainless steel rod with grooves for a NiCr heating wire with a diameter of 0.5\,mm. The wire is wound onto the rod with a bifilar winding, to minimize the magnetic field that it produces. This is important, because the main mirrors of GEO\,600 are suspended as triple pendulums and the upper stage uses coil magnet actuators for damping and alignment. Magnetic fields may couple into those actuators.

In addition to the direct radiation, the side heaters also use a polished aluminium reflector. 
Fig. \ref{fig:Assembly4-only-for-presentation-with-labels} shows a drawing of one of the side heaters. The side heaters have been installed at a distance of about 10\,cm from the mirror. 

\section{Side Heater Performance}

\begin{figure}
	\centering
		\includegraphics[width=1.00\textwidth]{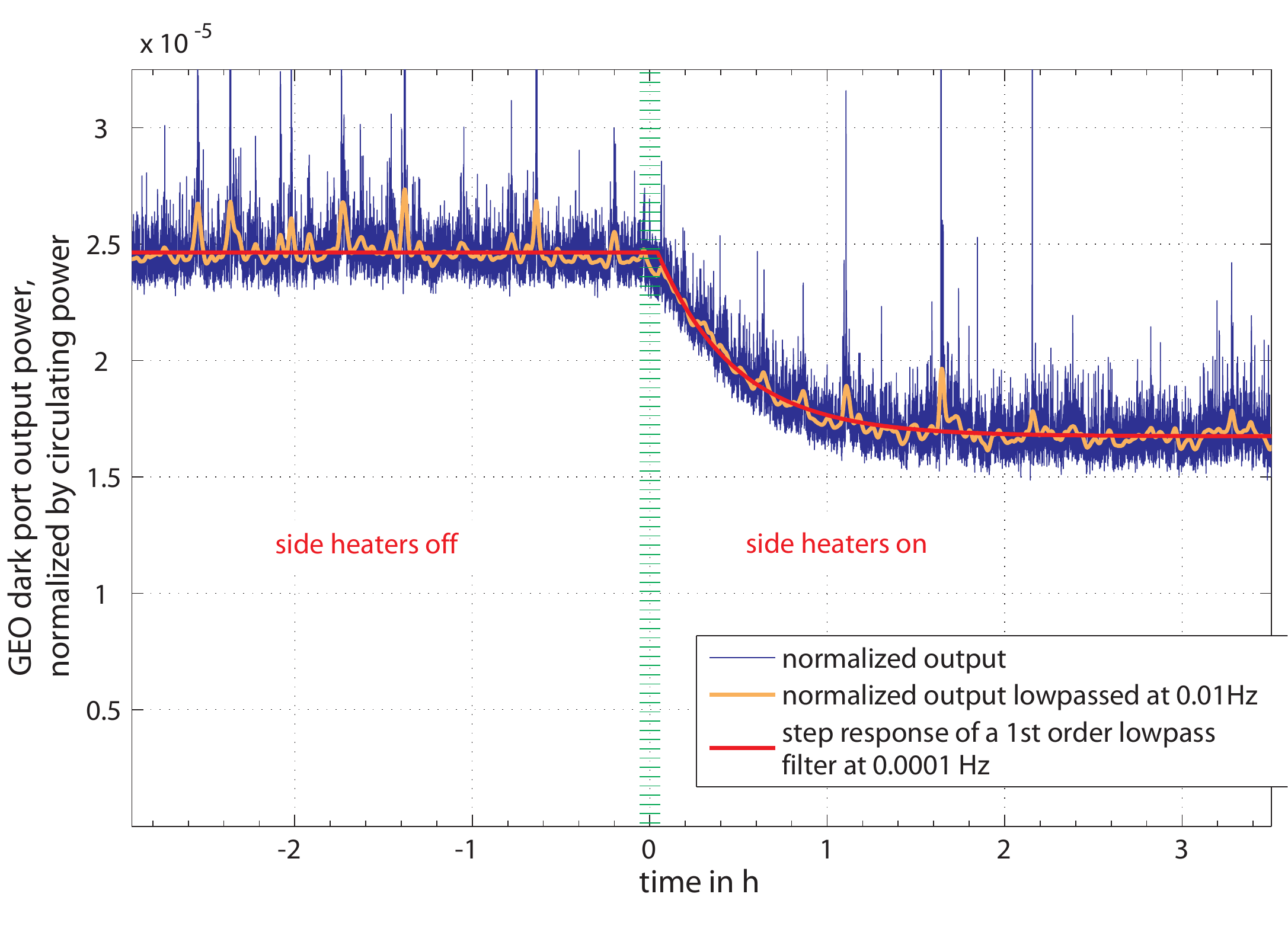}
	\caption{Time series of the  normalized power leaving the dark port of GEO\,600. At the start of the experiment the side heaters are off. They are turned on at the time marked as zero hours on the time axis.}
	\label{fig:quadvis-neu}
\end{figure}

We find that the side heaters are a good tool to reduce the HOM content at the dark port of GEO\,600. For a circulating light power of about 2.5\,kW in the PRC and a ring heater power giving the lowest power at the output, the optimal power of the side heaters is 2.5\,W each. Subsequent optimization of the ring heater power shows that this is the optimum setting. We can reduce the total power at the output of GEO\,600, which is dominated by HOMs, from 55\,mW to 37\,mW. Since the TEM$_{00}$ light has a power of 6\,mW at the dark port of GEO\,600, the side heaters reduce the power of HOMs in GEO\,600 by about 37\%.

\begin{figure}%
\centering
\includegraphics [width=1\textwidth] {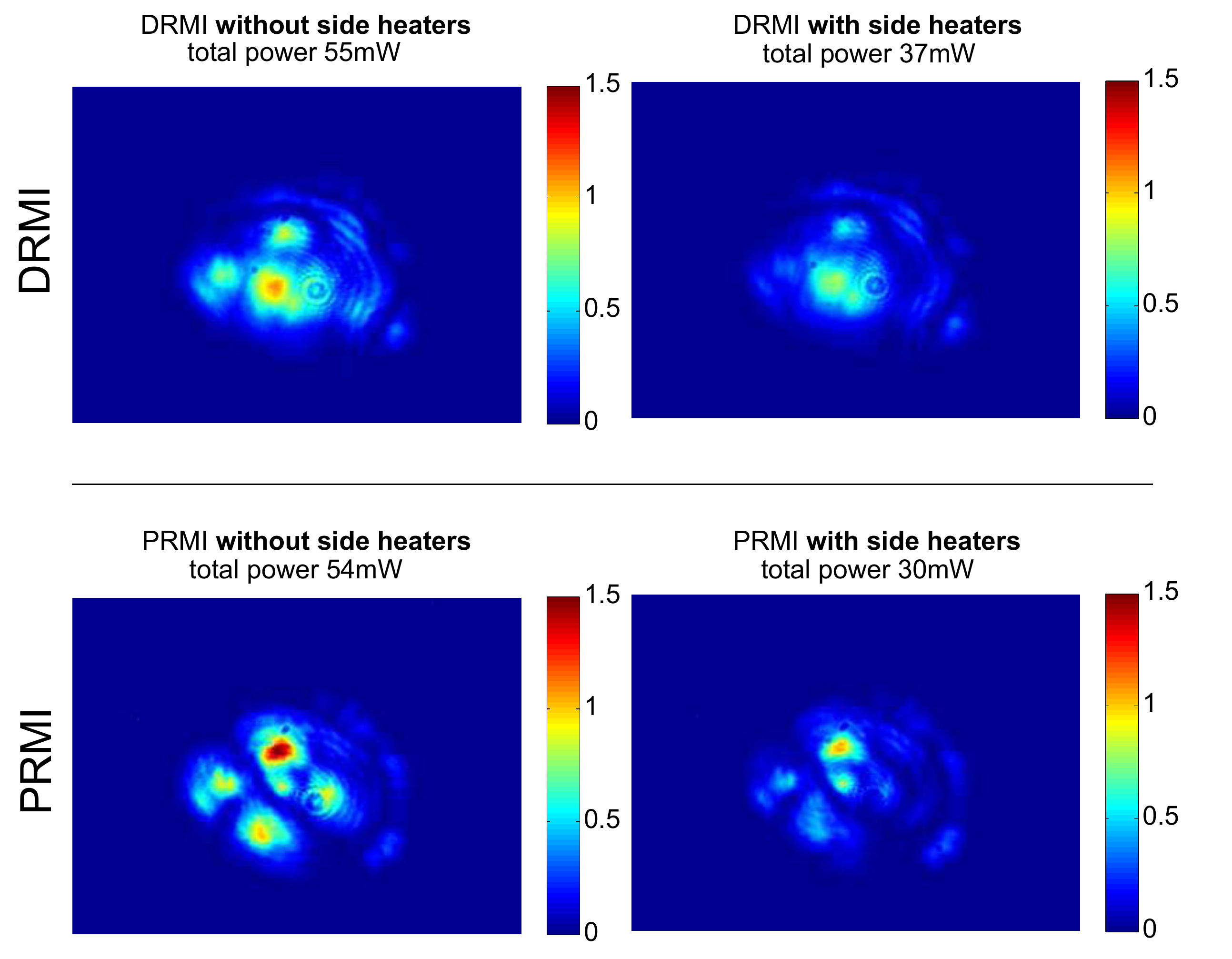}
\caption{This figure shows calibrated (to Watt per total image area) CCD images of the dark port output of GEO\,600. The top row shows the dark port image in normal (DRMI) operation, with and without the side heaters. The bottom row shows the situation in PRMI mode. Especially in PRMI mode, but also in DRMI mode, the lower power in HOMs is apparent.}%
\label{fig:drmi-comp}%
\end{figure}

In order to exclude any effect that the side heaters may have on the amount of circulating light power, we also look at the total ouput power normalized by the circulating power. This normalized output power decreases by 30\% with the side heaters. A plot of the normalized total output power is provided in figure \ref{fig:quadvis-neu}, while figure \ref{fig:drmi-comp} shows CCD camera images of the output beam without and with side heaters, respectively. The time constant of the side heaters can be fitted with a first order lowpass filter with a cutoff frequency of $f= 10^{-4}$\,Hz, which gives the time constant of $t=1/f \approx$ 2.8\,hours.

GEO\,600 can also be operated without signal recycling, as a power-recycled Michelson interferometer (PRMI). In that operating mode, the effect of the side heaters is even more pronounced because there is no mode healing effect of the signal recycling cavity. In PRMI mode, the side heaters reduce the power at the dark port from 54\,mW to 30\,mW. When normalized to the circulating light power, this corresponds to a reduction of 50\,\%. In fact, when using the side heaters, the normalized output power of GEO\,600 in PRMI mode is the same as in the normal, dual-recycled, operating mode. This demonstrates that we can ideally match the east mirror to the north mirror in two degrees of freedom. This equal to a curvature change of the mirror from (665.5/658 $\pm$ 2)\,m without the side heaters to (661/661 $\pm$ 2)\,m with the side heaters at the ideal setting of 2.5\,W.

\section{Summary}
GEO\,600 uses a ring heater behind the far east mirror to correct its RoC in one degree of freedom. We find that this ring heater induces astigmatism at that mirror, which leads to an increase in higher order mode content at the output of the interferometer. We designed and installed additional heaters laterally at the same mirror to correct the astigmatism, and reduce the power in unwanted HOMs at the darkport by 37\%. Since their installation in November 2012, the side heaters have been running nearly continuously, with GEO\,600 taking astro-physical data.


\ack{We thank the GEO collaboration for the development and construction of 
GEO\,600. The authors are also grateful for support from
the Science and Technology Facilities Council (STFC),
the University of Glasgow in the UK,
the Max Planck Society,
the Bundesministerium f\"ur Bildung und Forschung (BMBF),
the Volkswagen Stiftung,
the cluster of excellence QUEST (Centre for Quantum Engineering and 
Space-Time
Research),
the international Max Planck Research School (IMPRS),
and the State of Niedersachsen in Germany.
}

\end{document}